# Effects of stimulation frequencies on energy efficiency of a muscle fiber during contraction

Jiaxiang Xu[1], Bin Chen [1,2]*

[1]Department of Engineering Mechanics, Zhejiang University, Hangzhou, China

[2]Key Laboratory of Soft Machines and Smart Devices of Zhejiang Province, Hangzhou, China

* To whom correspondence should be addressed: chenb6@zju.edu.cn

**Abstract**

Contradictory experimental reports on the relationship between efficiency and stimulation frequency have hindered mechanistic understanding in converting neural activity into mechanical work during muscle contraction. To resolve this issue, we develop a biophysical model integrating calcium-mediated excitation with a detailed cross-bridge cycle to enable single-fiber simulations. Our model predictions indicate that the emergent shortening velocity is the primary determinant of cross-bridge efficiency: efficiency peaks at an optimal velocity and declines at higher or lower velocities, while frequency appears to exert secondary influence. Critically, the velocity yielding peak efficiency remains almost consistent across frequencies, with a slight upward shift at higher frequencies in most of our parametric studies. Interestingly, elevated inorganic phosphate ([Pi]) appears to amplify the efficiency disparity between high- and low-frequency regimes in our analysis. Our work suggests that stimulation frequency modulates efficiency predominantly through its regulation of shortening velocity, which primarily governs the kinetics of the myosin power stroke. This work may help clarify neural control of muscle energetics, and provide a quantitative foundation for studying muscle function in physiological and pathological contexts.

## Secondary Abstract

Experimental studies report conflicting relationships between stimulation frequency and muscle efficiency. To help reconcile these observations, we developed a biophysical model that couples calcium dynamics with cross-bridge cycling. Our simulations suggest that shortening velocity is a major determinant of cross-bridge efficiency, while stimulation frequency plays an indirect role by modulating the number of active myosin motors. The velocity at which efficiency peaks appears similar across frequencies, with a modest upward shift at higher frequencies. Elevated inorganic phosphate, as occurs during fatigue, may widen the efficiency gap between high- and low-frequency contractions. These findings offer a quantitative framework for understanding how neural activation patterns influence muscle energetics under physiological and pathological conditions.

## Introduction

Skeletal muscle fulfills its fundamental role as an effector for movement by efficiently transducing electrical signals from the nervous system into mechanical work and motion (1). The performance of this transduction process determines an organism's motor capability and energy economy. Within the regulatory strategies of the nervous system, the firing rate of motor neurons, also termed as the stimulation frequency, is a key parameter for the fine control of the dynamic range of muscle force output (2-5). It is noteworthy, however, that muscle performance is not solely defined by the magnitude of force produced and the energy consumption or the efficiency of converting chemical energy into mechanical work is also a central metric for assessing neuromuscular system performance (6). Understanding how neural regulation influences muscle efficiency holds profound implications for elucidating the principles of motor control, developing rehabilitation strategies for neuromuscular diseases, and advancing related pharmaceutical research (7, 8).

Muscle efficiency is governed by intrinsic factors such as fiber-type composition, temperature, mechanical load, etc (9-11). While neural modulation of muscle activation is well characterized, its impact on efficiency remains debated. Early studies reported elevated efficiency in maximally activated insect flight muscle (12), whereas others observed reduced efficiency in fully recruited frog slow-twitch muscle fibers (13)—a discrepancy potentially attributable to unaccounted work performed by series elastic elements (14). Later studies either confirmed or contradicted these earlier observations (14, 15), highlighting persistent contradictions that may reflect methodological differences across experimental systems.

These discrepancies may also arise from divergent definitions of efficiency, particularly regarding inclusion criteria for energy cost. Classical measurements (16) aggregated energy expenditures from ancillary processes (e.g., ion pumping, mitochondrial respiration). Partly due to the pronounced nonlinearity (17), quantifying the full energy cost that is associated with action potential propagation and the subsequent cascade of biochemical events it triggers remains challenging (18), complicating the interpretation of overall energy efficiency. Modern frameworks partitioned efficiency into mitochondrial (near-constant) and cross-bridge (load-dependent) components (19). In this context, a mechanism-driven mathematical model is desirable—one capable of simulating the sequence of events from action-potential-triggered calcium release and reuptake, through the actomyosin cross-bridge cycle, to active force generation. Such a model would allow the isolated analysis of how stimulation frequency regulates contraction energetics of individual muscle fiber, which can be compared with experiment and independent of confounding whole-muscle effects.

Previously, we developed a mechanics model of the sarcomere unit (20), which successfully reproduced a wide spectrum of experimental observations, including the effects of both constant [Pi] and rapid [Pi] jumps on contraction dynamics, demonstrating robust generalization capability. Here, we combine that mechanics model with the $Ca^{2+}$ kinetics model (21) to probe the synergistic effects of calcium transients on muscle contraction under neural regulation. Compared to other previous models (22-28), in our own opinion, our current model may provide a more physiologically realistic representation of myosin dynamics across multiple nucleotide states, the energy conversion process, and its sensitivity to [Pi]. Consequently, it may offer a more reliable theoretical foundation for elucidating the mechanisms of neuro-muscular coupling. By analyzing the mechanical behavior of muscle during active contraction under different stimulation frequencies, this work may help elucidate the impact of stimulation frequency on skeletal muscle contraction efficiency, which also creates a quantitative basis for the analysis of muscle function in physiological and pathological states.

## Method

a) Model

The basic functional unit during muscle contraction is a motor unit, which generally consists of a motor neuron and muscle fibers it innervates. When a motor neuron is excited and its firing frequency reaches the threshold for a motor unit (29), a generated action potential propagates along the sarcolemma via the T-tubules, which regulates $Ca^{2+}$ dynamics and down-stream muscle fibers contraction within the motor unit (30), as illustrated in Fig. 1. In this work, we consider that there already exists a generated action potential propagating along the T-tubule membrane with a resting potential denoted as $V_0$. T-tubule membrane potential, denoted as $V$, is varied solely in its stimulation frequency while its waveform and amplitude, denoted as $V_a$, are fixed (Please see Supplemental Information).

The dihydropyridine receptors (DHPRs) (31) on T-tubule membrane are composed of four subunits (32). Each DHPR is mechanically coupled to a ryanodine receptor (RyR) on the sarcoplasmic reticulum (SR) membrane (33), which is a calcium channel protein located on the SR membrane (30). Following Rios et al. (31) as described here, we consider a regulatory complex comprised of a DHPR and a RyR. All four voltage sensors of a DHPR are initially at the resting state (31). The initial opening rate of a RyR channel is denoted as $k_R$, and its initial closing rate $k_{-R}$. When $m$ voltage sensors of a DHPR are activated, the RyR channel opening rate increases to $\frac{k_R}{\gamma^m}$, while the closing rate decreases to $\gamma^m k_{-R}$, where $\gamma$ is a constant less than 1 (31). Vice versa, the rate at which voltage sensors on a DHPR are activated also depends on the state of the RyR channel (31). For a closed RyR channel, the rate for the transition of each voltage sensor of DHPR from the resting state to the activated state is denoted as $k_D$, and the corresponding backward rate is denoted as $k_{-D}$. When the RyR channel is at the open state, the activation rate of each voltage sensor of DHPR increases to $\frac{k_D}{\gamma}$, while its deactivation rate decreases to $\gamma k_{-D}$ (31). With $k_R$, $k_{-R}$, $k_D$ and $k_D$, the proportion of open RyR calcium channels on the T-tubule membrane at any time due to the change in T-tubule membrane potential, which is denoted as $r$, can be calculated.

Adopting the two-compartment modeling approach (21, 34) as described here, we consider that there exist four regions with different calcium ion concentrations, where the SR is partitioned into the terminal sarcoplasmic reticulum (TSR) and the general sarcoplasmic reticulum (GSR), while the myoplasm is divided into the subspace adjacent to the TSR (TMP) and the bulk myoplasm (BMP). For clarity, TSR, GSR, TMP, and BMP will be employed as subscripts in notation to refer to these respective regions.

Molecules including $Ca^{2+}$, $Mg^{2+}$, or their formed complexes with ATP can diffuse between GSR and TSR or between BMP and TMP (21, 34). The intercompartmental diffusion of calcium ions is described by $\frac{d[Ca_s]}{dt} = -\tau_{Ca}\frac{[Ca_s]-[Ca_d]}{Vol_s}$ or $\frac{d[Ca_d]}{dt} = \tau_{Ca}\frac{[Ca_s]-[Ca_d]}{Vol_d}$, where $\tau_{Ca}$ is the intercompartmental $Ca^{2+}$ diffusion parameter, $[Ca_s]$ and $[Ca_d]$ represent the calcium ion concentrations in the source and destination compartments, respectively, and $Vol_s$ and $Vol_d$ denote the volumes of the respective compartments (35, 36). The diffusion of $Mg^{2+}$ can be described by similar equations but with a different intercompartmental $Mg^{2+}$ diffusion parameter, $\tau_{Mg}$. The diffusion of $Ca^{2+}$ or $Mg^{2+}$ ions complexed with ATP is governed by the similar equations but with the intercompartmental Ca·ATP diffusion parameter and the intercompartmental Mg·ATP diffusion parameter, which are assumed to be the same and represented by $\tau_{ATP}$ (21).

Transmembrane ion transport is described by a different diffusion equation. TSR releases most of the calcium ions (37) to TMP through open RyR channels, described by $\frac{d[Ca_{TMP}]}{dt} = k_a r\frac{[Ca_{TSR}]-[Ca_{TMP}]}{Vol_{TMP}}$, where $k_a$ is a constant and $Vol_{TMP}$ is the volume of TMP (21, 31). Calcium ions can also leak directly across SR membrane into MP, described by $\frac{d[Ca_{TMP}]}{dt} = k_p \frac{[Ca_{TSR}]-[Ca_{TMP}]}{Vol_{TMP}}$ and $\frac{d[Ca_{BMP}]}{dt} = k_p \frac{[Ca_{GSR}]-[Ca_{BMP}]}{Vol_{BMP}}$, where $k_p$ is a constant (21). On the other hand, calcium ions in BMP are actively pumped back into the SR compartment by SERCA, an ATP-consuming pump protein located on the SR membrane (38), described by $\frac{d[Ca_{GSR}]}{dt} = k_m \frac{[Ca_{BMP}]}{C_0+[Ca_{BMP}]}/Vol_{GSR}$ and $\frac{d[Ca_{TSR}]}{dt} = k_m \frac{[Ca_{TMP}]}{C_0+[Ca_{TMP}]}/Vol_{TSR}$, where $k_m$ is the maximum rate, $C_0$ is a constant, $Vol_{GSR}$ is the volume of GSR and $Vol_{TSR}$ is the volume of TSR (39).

The reaction rates of calcium with proteins, including parvalbumin, which, however, are missing in slow muscle fibers, calsequestrin, troponin, or ATP in the two-compartment model was described using the standard mass-action rate law for the first-order or pseudo-first-order processes (21, 36). The complete set of reaction equations possess the general form but are characterized with distinct rate constants, which are provided in the Supplemental Information.

The state transition diagram for the processes of calcium regulation of the interaction between actin and myosin motors (21, 40, 41) is depicted in Fig. 1b. The regulatory unit is comprised of troponin, tropomyosin, actin and myosin. The rate constant for a calcium binding, denoted as $k_{Ca}$, or that for the release of a single calcium ion, denoted as $k_{-Ca}$, is identical no matter whether the troponin is bound with or without calcium ion (21). The exposure process of tropomyosin is simplified in the model with an exposure rate constant, denoted as $k_e$. Only when the troponin is bound with two $Ca^{2+}$ ions and actin has its binding sites exposed, the actin can interact with myosin motors (42). When a myosin motor is attached to the actin, the actin cannot revert to the tropomyosin-blocked state, nor can it release its bound calcium ions. As depicted in Fig. 1b, for the reverse transition to the tropomyosin-blocked state, the transition rate constant from the state of the troponin bound with two $Ca^{2+}$ ions is denoted as $k_{b2}$, while that from the state of the troponin bound with none or only one $Ca^{2+}$ ion is denoted as $k_{b1}$.

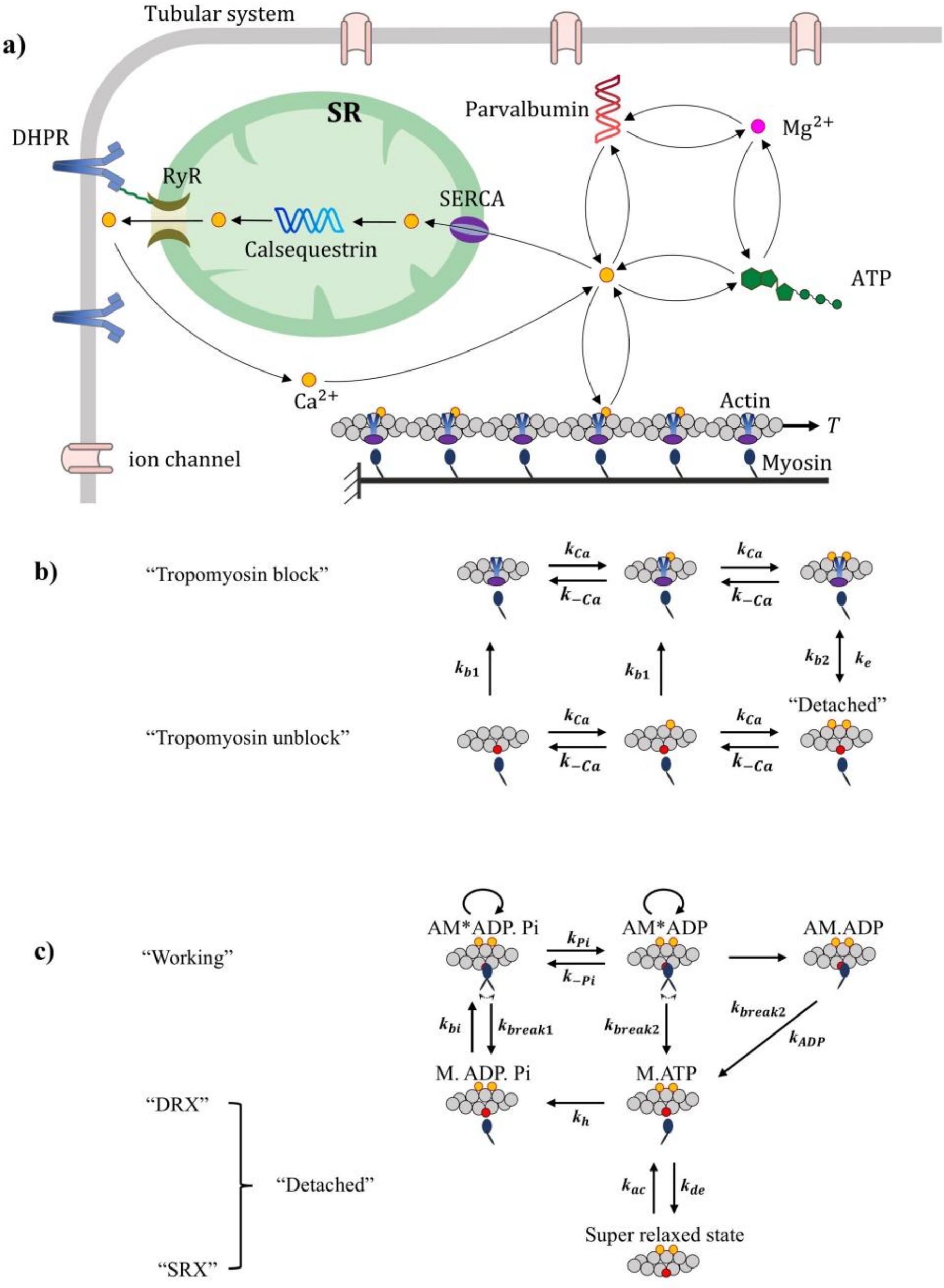

Fig. 1 a) Schematic of excitation-contraction coupling in a muscle fiber. The T-tubule membrane, maintained by ion pumps, conducts action potentials and activates DHPRs, which mechanically couple to RyRs on SR. Stored $Ca^{2+}$ are subsequently released into the myoplasm. There, free $Ca^{2+}$ may bind to ATP or parvalbumin, with $Mg^{2+}$ competing for the same sites. The remaining $Ca^{2+}$ diffuses to the sarcomere and binds to troponin, initiating fiber activation. b) State map of the Regulatory Unit of actin activation (21): Initially, tropomyosin blocks the myosin-binding sites on actin. Binding of two $Ca^{2+}$ ions to troponin shifts tropomyosin to an unblocked state, exposing the actin sites for myosin binding. Loss of $Ca^{2+}$ ions causes tropomyosin to shift back, restoring the blocked state. c) The chemomechanical cycle of the myosin cross-bridge (20): The M·ATP, M·ADP·$P_i$ and SRX states are detached. Myosin may enter the SRX state from M·ATP. Binding to an unblocked actin site transitions myosin into AM*ADP·Pi. $P_i$ release (AM*ADP) and the power stroke are uncoupled. Following ADP release at the AM·ADP state, myosin detaches after securing a new ATP. Premature detachment from working states of AM*ADP·Pi and AM*ADP can also occur via forced rupture. A motor detached from AM*ADP.Pi can rebind to the actin to perform more swings.

Table 1 Default values of parameters for calcium cycle dynamics used in the simulations

| Parameter | Value | Parameter | Value |
|---|---|---|---|
| $V_0$ | -80mV (21) | $V_a$ | 35mV (21) |
| $k_R$ | 0.002 /ms (31) | $k_{-R}$ | 900 /ms (31) |
| $\gamma$ | 0.175 (31) | $k_{b2}$ | 0.15 /ms (31) |
| $k_{\mathrm{a}}$ | 60 $\mu m^3 \cdot ms^{-1}$ (31) | $k_p$ | 0.00004 $\mu m^3 \cdot ms^{-1}$ (21) |
| $k_{\mathrm{m}}$ | 2.4 $\mu M \cdot \mu m^{-3} \cdot ms^{-1}$ (21) | $c_0$ | 1 $\mu M$ (36) |
| $\tau_{Ca}$ | 0.75 $\mu m^3 \cdot ms^{-1}$ (21) | $\tau_{Mg}$ | 1.5 $\mu m^3 \cdot ms^{-1}$ (21) |
| $\tau_{ATP}$ | 0.375 $\mu m^3 \cdot ms^{-1}$ (21) | $k_e$ | 0.15 $ms^{-1}$ (21) |
| $k_{b1}$ | 0.05 $\cdot ms^{-1}$ (21) | $k_{b2}$ | 0.15 $ms^{-1}$ (21) |
| $Vol_{TSR}$ | 0.0004 $\mu m^3$ (21) | $Vol_{GSR}$ | 0.04 $\mu m^3$ (21) |
| $Vol_{TMP}$ | 0.008 $\mu m^3$ (21) | $Vol_{BMP}$ | 0.8 $\mu m^3$ (21) |

Following our previous work (20) as described here, we employ a mechanics model to describe the contraction dynamics of muscle fibers due to interactions between individual myosin motors and their binding sites on actin filaments. As depicted in Fig. 1a, we consider only a quarter of the sarcomere, which is the basic structural unit of a muscle fiber and comprises of an elastic thin filament (axial rigidity $EA_{\mathrm{thin}}$) and a thick filament (axial rigidity $EA_{\text{thick}}$), the latter featuring $N_m$ uniformly distributed myosin motors spaced at intervals $l_m$. The thick filament is fixed at its left end, while the thin filament experiences an applied filament load $T$ at its right end.

As shown in Fig. 1c, the chemomechanical cycle of a single myosin motor is modeled through six states, primarily following the Lymn-Taylor theory (43). The cycle begins with the super-relaxed (SRX) state (44), where myosin cannot bind actin. Transition rates between SRX and M·ATP are load-dependent (45). ATP hydrolysis drives the shift from M·ATP to M·ADP·Pi, enabling actin binding via Brownian motion. Upon binding, myosin enters AM*ADP·Pi and then AM*ADP (post-Pi release), with Pi-release uncoupled from the lever-arm swing (20). At these states, the lever arm executes multiple forward swings (46), each with force-dependent kinetics: the swing rate $k_s = k_{s0} \exp\left(\frac{f_a - f}{f_a}\right)$ (where $k_{s0}$ is the rate at $f = f_a$) proceeds only if motor force $f < f_a$, while $f > f_a$ arrests swinging (backward swings are neglected). The swing distance for each step, $\Delta s = \frac{f_a - f}{q}$ ($q$: motor spring constant), accumulates until reaching a maximum $L_s$, after which ADP is released and ATP rebinding triggers rapid detachment. Alternatively, forced detachment at AM*ADP·Pi allows rebinding with retained ADP and Pi (47), while detachment at AM*ADP or AM·ADP involves losing ADP and quickly capturing ATP to return myosin to M·ATP. The total swing distance per hydrolyzed ATP is capped at $L_s$, irrespective of rebinding events (20).

The mechanics model (18) accounts for significant transition rates between myosin states (Fig. 1c), where the SRX-to-M·ATP transition follows Bell's formula (48) as $k_{ac} = k_{01} \exp\left(\frac{T}{T_{01}}\right)$ with $k_{01}$ being a rate constant, while the reverse transition $k_{de} = k_{10} \exp\left(-\frac{T}{T_{01}}\right)$ with $k_{10}$ being a rate constant. ATP hydrolysis proceeds at a constant rate $k_h$. Myosin is modeled as a linear spring coupled to an active lever arm, with its actin-binding rate $k_{bi}$ involving myosin spring constant $q$ and binding site spacing $l_a$. Pi release $k_{Pi}$ is constant, while rebinding $k_{-Pi}$ is disabled when Pi concentration is low by default. Force-generation follows Hookes' law $f = qx$, with stretch $x = u + d$, where $u$ is the separation between the myosin and its binding site on the thin filament and $d$ is the accumulated swing distance of the lever arm. The myosin-actin bond exhibits catch-slip behavior, where detachment rates differ between states: $\mathrm{k_{break1}}$ for AM*ADP·Pi, and $\mathrm{k_{break2}}$ for AM*ADP (20). The cycle concludes via a fixed-rate $k_{ADP}$ for ADP release and ATP rebinding. Initially, $N_a$ constitutively active motors are in the AM*ADP·Pi state (45), while the remaining motors begin in the SRX state.

b) Numerical scheme

A coupled Monte Carlo method and finite element method (49) is employed to simulate the model (20, 50-51). The evolution trajectory of the proportion of activated binding sites on actin filaments is solved with a set of differential equations, which directly modulates the motor attachment rates. The finite element method is used to calculate the force and displacement of each myosin motor, where thin and thick filaments within the half-sarcomere are discretized using one-dimensional two-node rod elements, while myosin motors are represented as one-dimensional spring elements. The First Reaction Method is employed to determine both the position and timing of next random event (52, 53). After calculating the occurrence time, $\Delta t_j$, for all possible random jumps, the next time step, $\Delta t$, is selected as the minimum among all $\Delta t_j$. Upon completing this time step, the system is updated. Note that the current proportion of Regulatory Units in the “Detached” state to all Regulatory Units will be updated, which will be multiplied by the concentration of Regulatory Units to obtain the concentration of Regulatory Units in the “Detached” state in order to simulate the calcium kinetics at the next time step.

The software used in this study is self-coded. Default values of parameters for calcium kinetics used in the simulations are listed in Table 1. Default values of parameters for sarcomere contraction dynamics used in the simulations are listed in Table 2. Given the parameters listed in Table 2, the muscle fibers in our simulation could correspond to slow-twitch skeletal muscle fibers. Model validation is also performed, as given in Supplemental Information.

Table 2 Default values of parameters for sarcomere dynamics used in the simulations

| Parameter | Value | Parameter | Value |
|---|---|---|---|
| $N_m$ | 76 (54) | $N_a$ | 5 (20) |
| $EA_{\text{thick}}$ | 46 nN (20) | $EA_{\text{thin}}$ | 23 nN (20) |
| $l_m$ | 14.3 nm (55) | $l_a$ | 5.5 nm (56) |
| $\xi$ | 200 /s (20) | $T_{01}$ | 40 pN (20) |
| $k_{s0}$ | 1700 /s (54) | $q$ | 2 pN/nm (20) |
| $f_a$ | 5 pN (20) | $L_s$ | 6 nm (57) |
| $k_{01}$ | 200 /s (20) | $k_{10}$ | 1000 /s (20) |
| $k_{\text{ADP}}$ | 300 /s (20) | $k_h$ | 150 /s (20) |
| $k_{Pi}$ | 250 /s (58) | $k_{-Pi}$ | 0 /s (20) |
| $p_0$ | 83 pN.nm (59) | $L_T$ | 1000 nm |

When studying muscle fiber's single and multiple twitches or tetanic contraction, the simulation begins with the muscle fiber under isometric conditions while the T-tubule membrane is at the resting potential. For twitches, an appropriate number of action potentials with certain frequency are input to the T-tubule membrane for simulation, and this procedure is repeated 100 times. For tetanic contraction, a sustained train of action potentials with a specific frequency is applied until the muscle force reached its maximum. After maintaining this force at a plateau for a period, the membrane potential is returned to the resting state, and the simulation continues until the force declines to a stable baseline. This procedure is also repeated for 100 times.

During the simulation of steady-state muscle contraction, the calcium kinetic cycle upon a given stimulation frequency is firstly simulated. After 0.5 second of stimulation, the proportion of activated binding sites reaches a relatively stable value. The muscle is then loaded for an additional 0.5 second to obtain the displacement curve and other physical quantities during the loading phase. To reduce transient artifacts, the load is increased at a rate of 5000 pN/s. With the method given in Xu et al. (20), we subsequently calculate the shortening velocity versus the filament load, the output power versus the filament load, the power consumption versus the filament load, the energy efficiency versus the filament load, the energy efficiency versus the shortening velocity, etc. In investigating the effects of motor size, it is assumed that all muscle fibers within a motor unit behave the same. This simplification enables the calculation of the total load of the motor unit by scaling the sarcomere values with the number of muscle fibers within the motor unit.

## Results

a) Effect of stimulation frequencies

We begin by simulating a motor unit with a single muscle fiber. The simulated effects of stimulation frequencies on $Ca^{2+}$ dynamics are displayed in Figs. 2a-c, with corresponding prescribed T-tubular membrane potentials of different stimulation frequencies displayed in Fig. 2d. Figure 2a illustrates the temporal variation of the weighted average of the $Ca^{2+}$ concentrations in the TMP and BMP compartments, denoted as $Ca_{MP}$, under different stimulation frequencies, revealing that $Ca_{MP}$ oscillates in response to action potentials. When the action potential frequency is sufficiently high, $Ca_{MP}$ progressively rises and eventually plateaus at a steady level. Figure 2b reveals that the proportion of active binding sites on actin filaments, denoted as $R_{active}$, increases over time until reaching saturation at fixed stimulation frequencies. The saturation levels rise with higher stimulation frequencies but eventually plateau, as seen in Fig. 2c. These results clearly indicate that, relatively low stimulation frequencies, results in low intracellular $Ca^{2+}$ concentration, leading to a reduced proportion of activated binding sites on the thin filament, which will affect cross-bridge formation and the subsequent cross-bridge dynamics.

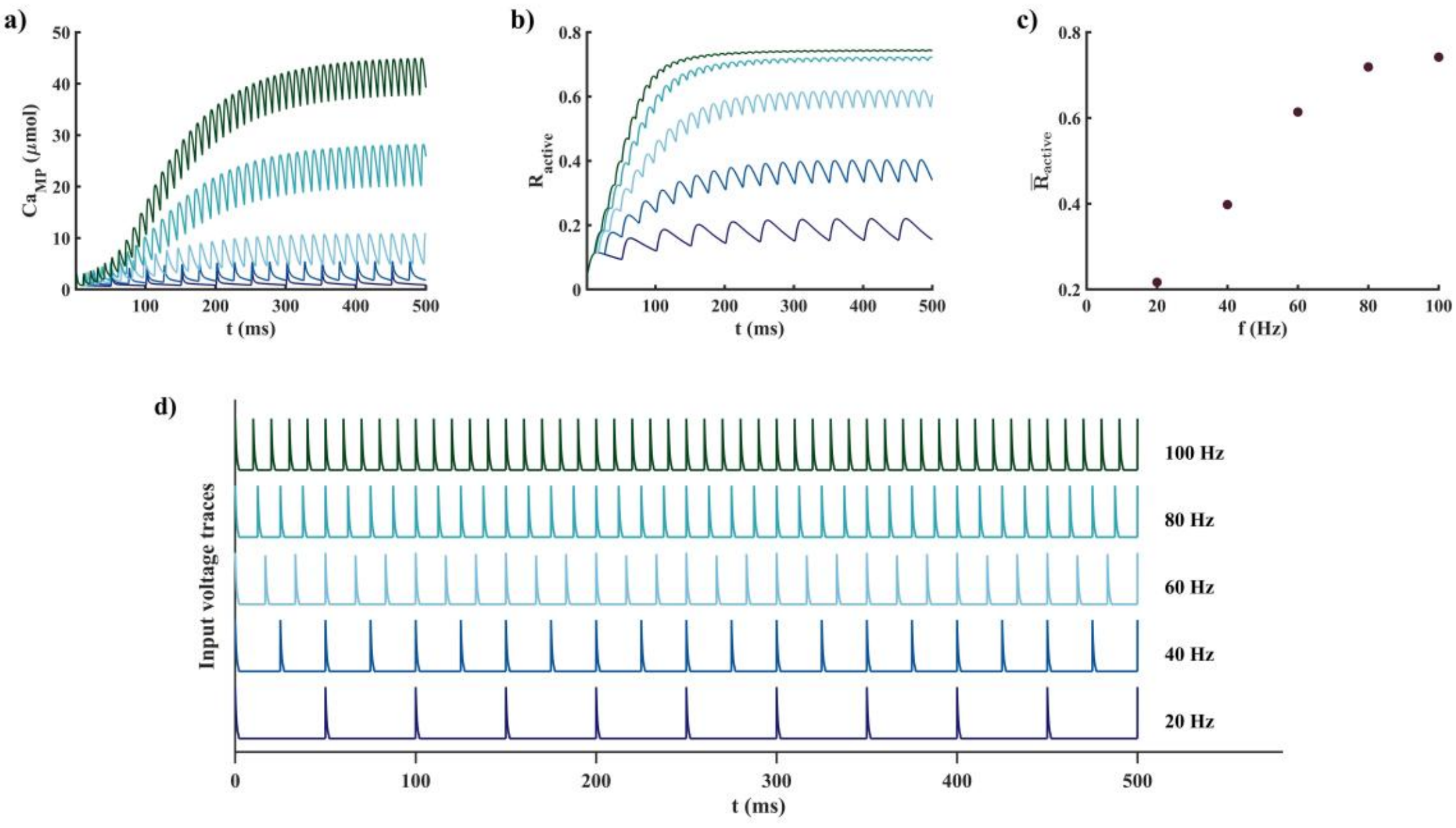

**Fig. 2** a) Time evolution of $Ca_{MP}$ in the cytoplasm; b) Time evolution of $R_{active}$ on the thin filament; c) The saturation value of $R_{active}$, denoted as $\bar{R}_{active}$, against stimulation frequency; d) Prescribed membrane action potential with different stimulation frequencies.

Predicted muscle contraction features across different stimulation frequencies are displayed in Fig. 3. For consistency, all filament loads are normalized by a reference load, $T_0$ (250 pN), close to the isometric load at 40 Hz (255 pN). All shortening velocities are normalized by a reference velocity $V_0 = L_T$ /s, which is 1000nm/s, where $L_T$ denotes the original length of the half sarcomere. The normalized filament loads and shortening velocities are denoted as TB and VB, respectively. Figure 3a displays the predicted force-velocity relation across a range of stimulation frequencies. A key observation from Fig. 3a is the reduction in both isometric load and contraction velocity at low stimulation frequencies. Specifically, the isometric load under 40 Hz stimulation is close to $T_0$, whereas at 20 Hz it drops to be ~$0.4T_0$. This decline correlates with a lower fraction of activated binding sites at reduced frequencies (Fig. 2b). Our simulation results align qualitatively with prior experimental findings (60) in Fig. 3g which reported that higher stimulation frequencies led to greater maximum force, larger contraction velocity and higher energy consumption. It should be noted, however, that the compared experiments (60) were performed on whole muscle, limiting the comparison to general trends rather than quantitative agreement.

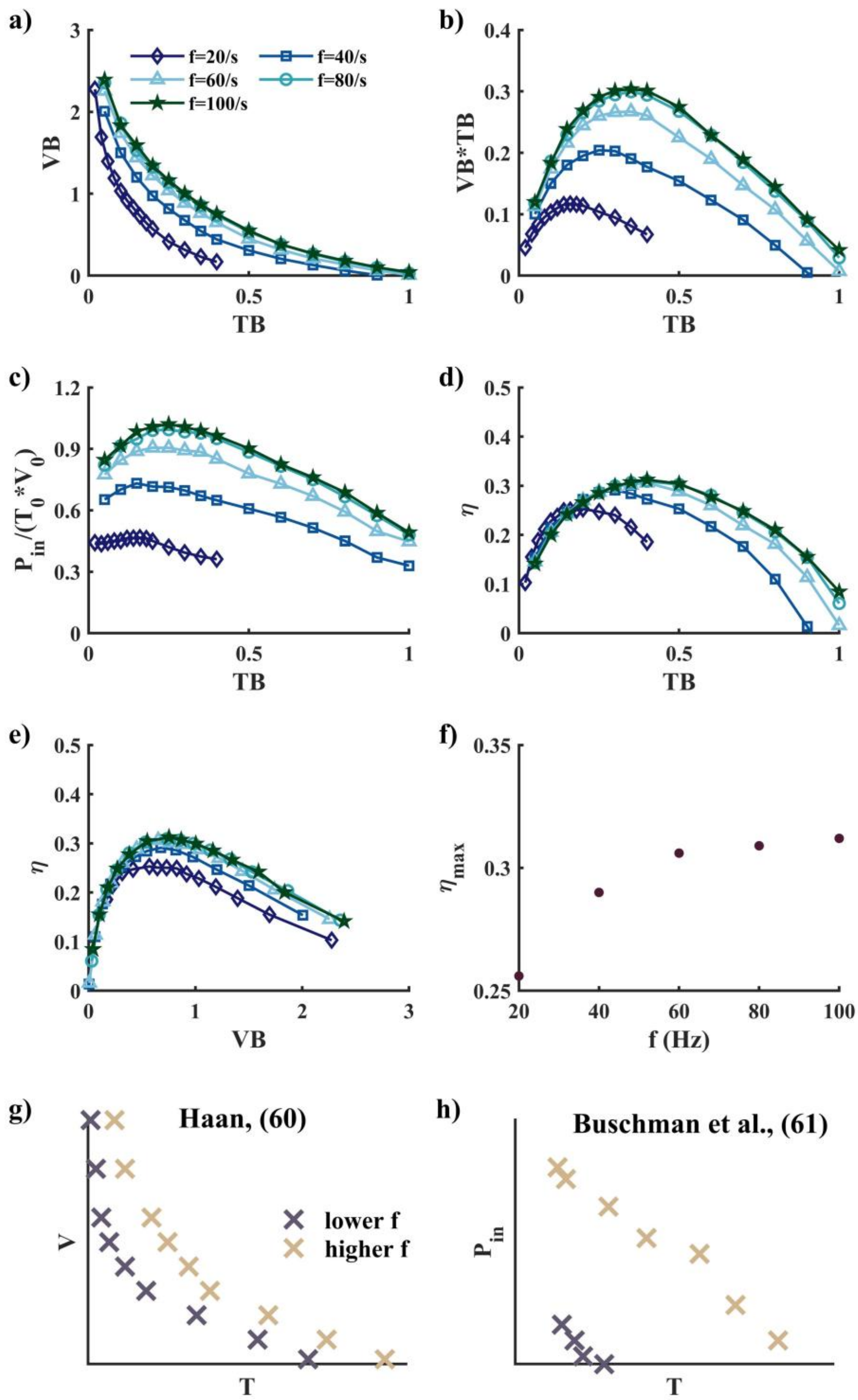

**Fig. 3** Predicted features of muscle contraction at different stimulation frequency (a-e): a) The shortening velocity versus the filament load; b) The output power versus the filament load; c) The power consumption versus the filament load; d) The energy efficiency versus the filament load; e) The energy efficiency versus the shortening velocity; f) Maximum value of the energy efficiency against stimulation frequency; (g) Reported shortening velocity versus the filament load from experiment (60); (h) Reported power consumption versus the filament load from experiments (61).

Figure 3b and 3c present the simulated output power and consumed power during active muscle contraction, respectively. Figure 3b shows that, across all stimulation frequencies, output power first rises and then declines with increasing load. Higher stimulation frequencies consistently produce greater output power under the same load, though this effect is more modest at lower loads. For example, at a load of $0.08T_0$, the output power at 80 Hz is approximately 1.22 times that at 40 Hz; at $0.2T_0$, this ratio increases to about 1.33, and at $0.4T_0$ it reaches 1.74. Power consumption likewise exhibits a rise-then-fall trend with increasing load (Fig. 3c), with higher stimulation frequencies leading to systematically greater energy consumption, while only a decreasing trend in power consumption as load increases can be inferred from a previous experiment (61), as displayed in Fig. 3h. Figure 3c also indicates a prominent Fenn effect at high activation frequencies that progressively diminishes at lower activation levels, where the rate of energy consumption becomes nearly independent of the load/shortening velocity. Taken together, these results indicate that while higher-frequency stimulation enables greater mechanical output, it also demands higher energy expenditure.

Figure 3d shows the energy efficiency ($\eta$) calculated from the results in Figs. 3b–c. The results show that, across all stimulation frequencies, efficiency first rises and then falls as load increases. At high loads, cross-bridge efficiency is greater under higher stimulation frequencies. In contrast, at low loads the differences between frequencies are very small and almost disappear. It should be noted that the efficiency presented here specifically represents cross-bridge efficiency, not overall thermodynamic efficiency.

Figure 3e replots the efficiency data from Fig. 3d as a function of shortening velocity,

as often reported in prior experimental studies. The plot shows that efficiency rises to a peak and then declines as shortening velocity increases. Notably, the shortening velocity at which peak efficiency occurs is similar across different stimulation frequencies. Figure 3e also indicates that higher stimulation frequencies generally yield slightly greater efficiency with the specific parameters adopted in our simulations. This effect, however, is not pronounced at very low shortening velocities; it becomes clearly evident within the intermediate or high velocity range. The peak efficiency under different stimulation frequencies is plotted separately in Fig. 3f, which increases with stimulation frequency and asymptotically approaches a plateau. As also seen from Fig. 3f, the difference in our simulated peak efficiency is pronounced at frequencies below 60 Hz, whereas the effect becomes relatively small above 60 Hz.

b) Effect of motor size

Motor units vary in size, and their recruitment is known to follow Henneman's size principle (29). To evaluate the effect of motor unit size, we linearly scale the simulated filament loads from Fig. 3 by its number of muscle fibers within a motor unit. The results are presented in Fig. 4. For normalization, the isometric load of a unit with 100 fibers under 100 Hz stimulation is defined as the reference value $T_1$. The normalized motor unit load by $T_1$ is denoted as TD. As shown in Figs. 4a and 4b, larger motor units sustain contraction over a broader load range. For example, at 40 Hz, a small unit (50 fibers) operates effectively only under loads up to $0.45T_1$, while a larger unit (100 fibers) functions across a range of $0\text{-}0.9T_1$. This indicates that recruiting larger units is necessary for greater force output. Output power increases with motor unit size under the same load, though larger units also consume more power under identical conditions (Figs. 4c-d).

Cross-bridge efficiency is also calculated (Figs. 4e-f). Under low loads, smaller units consistently show higher cross-bridge efficiency. In contrast, under high loads, larger units become more efficient. For intermediate loading conditions, an optimal size exists. For example, at $0.8T_1$ under 100 Hz stimulation, a unit with 200 fibers is optimal; slightly smaller (100 fibers) or larger (400 fibers) units are less efficient. The motor unit that achieves the highest efficiency depends on both load magnitude and stimulation frequency. Thus, the sequential recruitment of motor units by following Henneman's size principle (29) can allow muscle contraction to maintain relatively high efficiency across varying demands: smaller units are activated first to handle low loads efficiently, while larger units are engaged as needed to produce greater force, operating within their own high-efficiency range under higher loads.

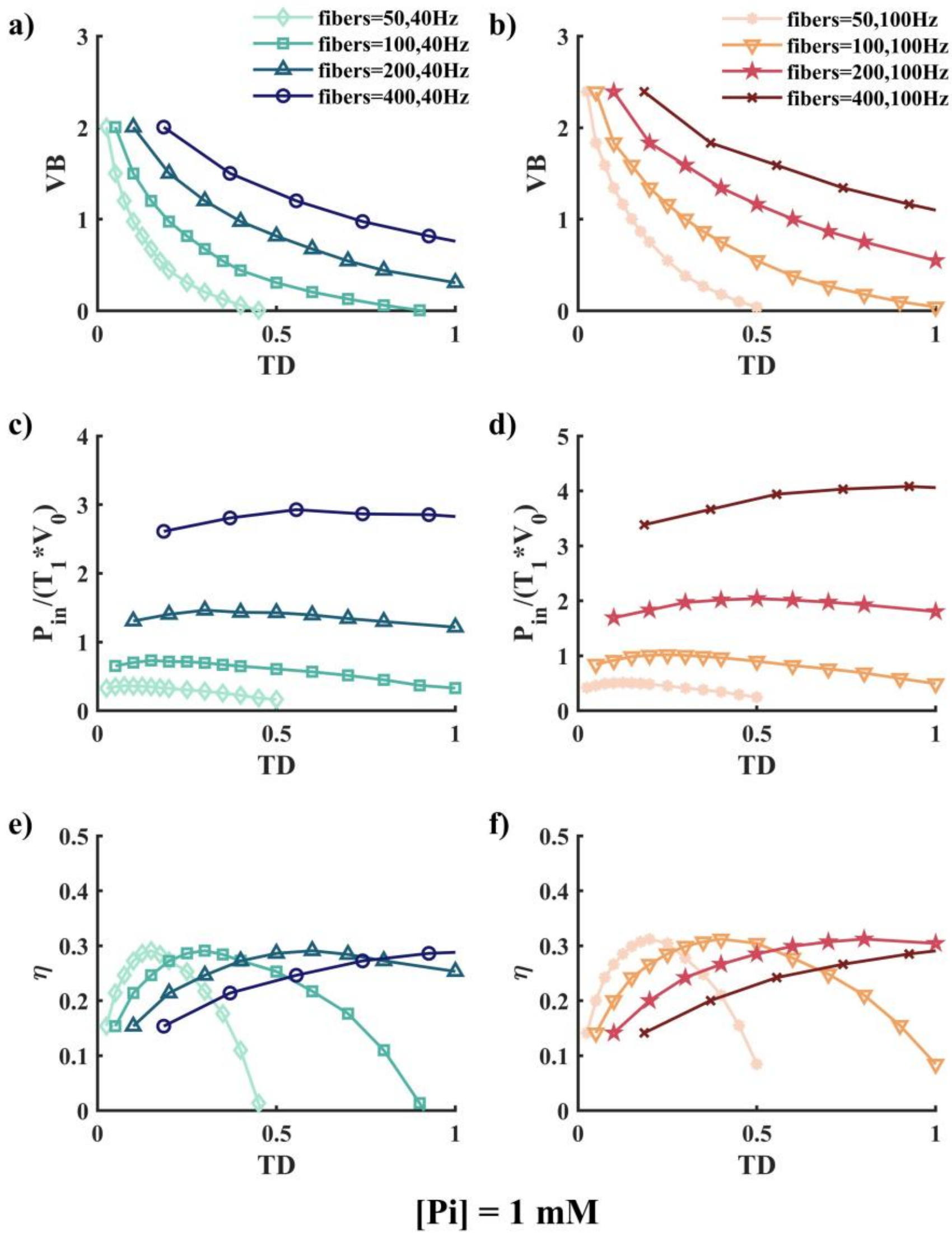

**Fig. 4** Predicted features of muscle contraction across varying motor unit sizes with a stimulation frequency of 40Hz (a, c, e) or 100Hz (b, d, f): (a, b) Shortening velocity as a function of motor unit load; (c, d) Power consumption vs. motor unit load; (e, f)

Energy efficiency vs. motor unit load.

c) Power stroke features of individual myosin motors

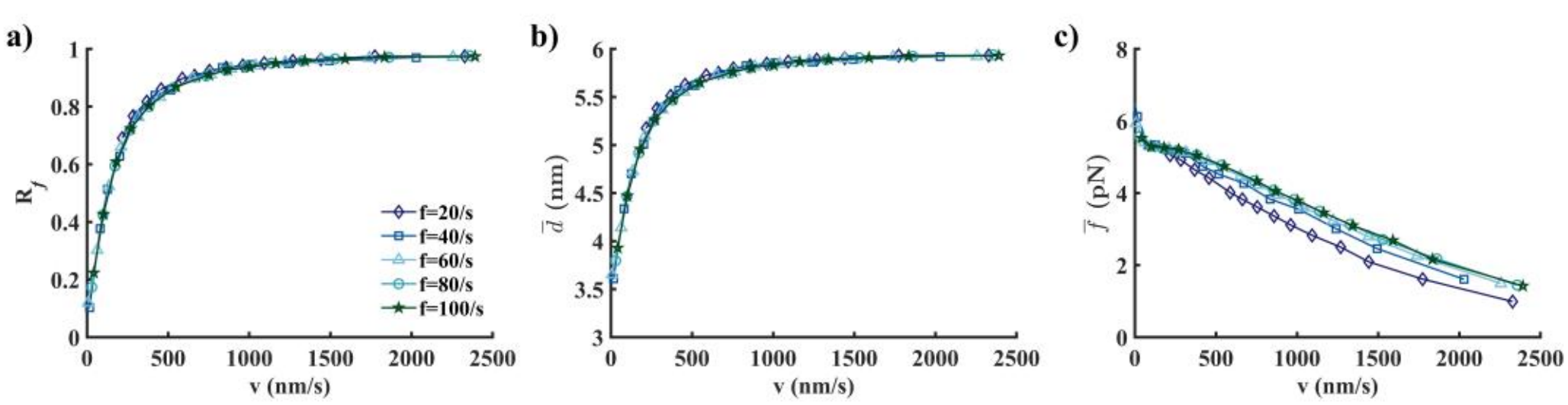

Fig. 5 a) The proportion of motors that execute full power stroke versus the shortening velocity; b) The average swing distance per motor versus the shortening velocity; c) The average motor force versus the shortening velocity.

Results in Fig. 3e show that cross-bridge efficiency is primarily governed by shortening velocity and is only secondarily modulated by stimulation frequency. In addition, the shortening velocity at which peak efficiency occurs remains remarkably robust across a broad range of stimulation frequencies and shows only a modest increase at the highest frequencies. To better understand these results, we furtherly examine how stimulation frequency influences the power stroke features of individual myosin motors. Using simulation outputs, we quantify the fraction of myosin motors that complete a full power stroke during active contraction, denoted as $R_f$, and plot this metric against shortening velocity in Fig. 5a. The data reveal that $R_f$ increases with shortening velocity and saturates near unity above ～750 nm/s, indicating that virtually all motors execute a complete power stroke when shortening exceeds this threshold. Crucially, the $R_f$ curves for different stimulation frequencies closely overlap, demonstrating that frequency affects $R_f$ predominantly through its regulating effect on shortening velocity. A more direct measure of power stroke efficacy than $R_f$ is the actual swing distance achieved per motor. Accordingly, we compute the mean power-stroke swing distance, denoted as $\bar{d}$, for all motors during shortening. When plotted against shortening velocity (Fig. 5b), $\bar{d}$ again yields nearly overlapping curves across stimulation frequencies.

We also calculate the average force generated by actively working motors during shortening, denoted as $\bar{f}$, shown in Fig. 5c. Once more, the force-velocity relationships are largely superimposable across frequencies. However, at identical shortening velocities, lower stimulation frequencies yield slightly reduced average motor forces.

The increase in peak efficiency with stimulation frequency in Fig. 3e apparently stems from the increase in the proportion of activated binding sites. A higher stimulation frequency raises $R_{active}$, which in turn increases the number of working myosin motors. This would alter the internal strain distribution along the thin filaments, which subsequently modifies the cross-bridge efficiency of individual motors. Considering that the average output work of a single motor due to energy released from a single ATP is approximately given by $\bar{f}$ multiplied by $\bar{d}$, the findings in Figs. 5b-c would account for what is observed in Fig. 3e. The difference in peak efficiency between different stimulation frequencies occurring at a shortening velocity of approximately 750nm/s is up to 20% in Fig. 3e. The corresponding difference in average motor force in Fig. 5c at the same shortening velocity is about 18%. Therefore, the modest increase in average motor force with stimulation frequency observed in Fig. 5c accounts for the frequency-dependent rise in peak efficiency seen in Fig. 3f. Collectively, the results in Fig. 5 support a mechanistic basis for our central conclusion: shortening velocity is the primary determinant of cross-bridge efficiency, while stimulation frequency exerts its influence mainly by modulating this velocity—thereby acting as a secondary regulator.

d) Effect of [Pi]

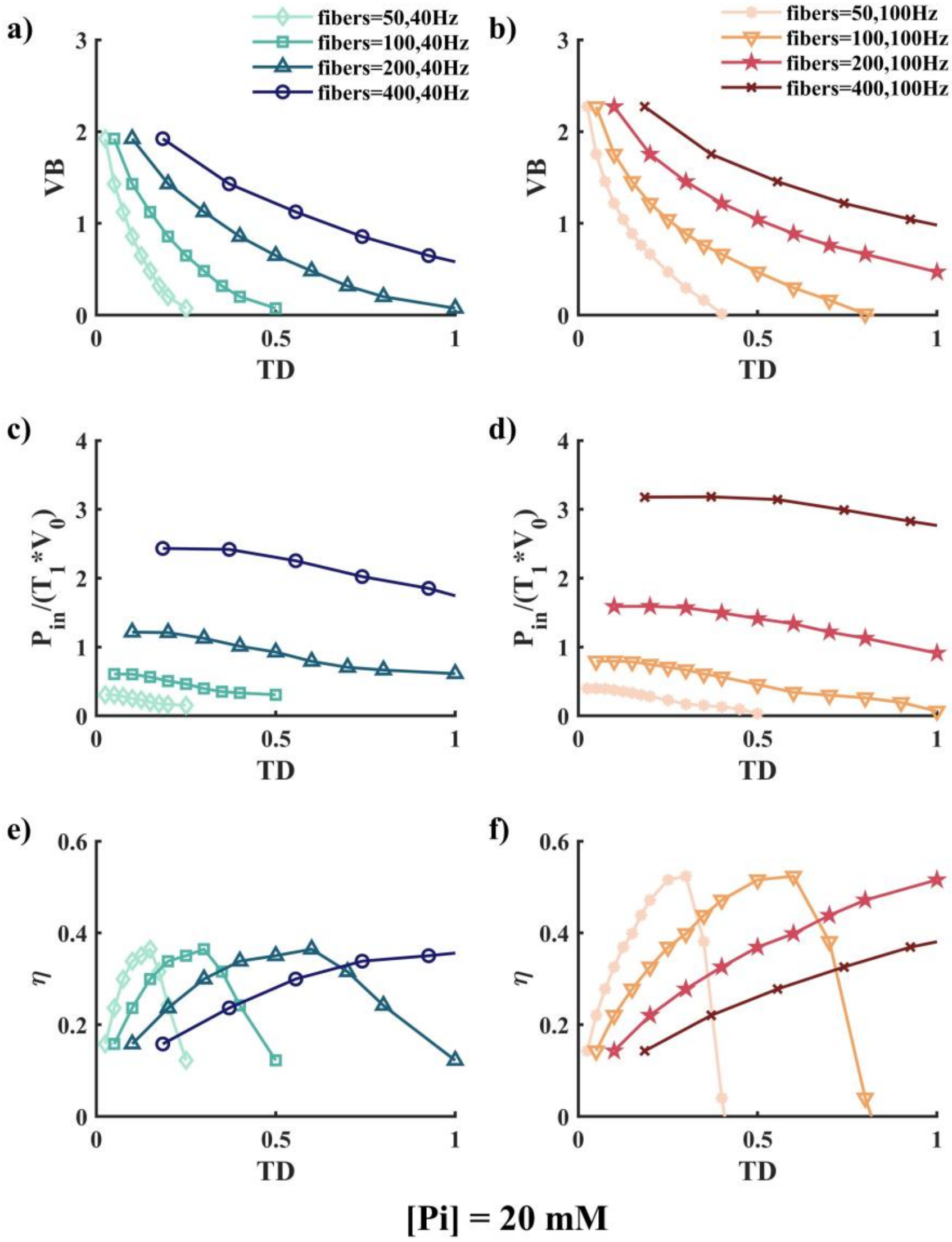

**Fig. 6** Predicted features of muscle contraction at [Pi]=20mM with a stimulation frequency of 40Hz (a, c, e) or 100Hz (b, d, f): (a, b) Shortening velocity as a function of motor unit load; (c, d) Power consumption vs. motor unit load; (e, f) Energy efficiency vs. motor unit load.

During muscle fatigue, the accumulation of inorganic phosphate (Pi) significantly alters cross-bridge cycling. To simulate this effect on motor unit contraction in our model, elevated [Pi] is represented by a higher rate of phosphate rebinding, $k_{-Pi}$, given by $k_{-Pi} = k_{max}[\mathrm{Pi}]/(c_M + [\mathrm{Pi}])$ (20), with $k_{max}$ being the maximal rebinding rate of Pi and $c_M$ the corresponding Michaelis constant (62). The motor unit behaviors under elevated [Pi] (20 mM) are shown in Fig. 6 for different stimulation frequencies and motor sizes, compared to a default of 1 mM in Fig. 4. One outcome is the significant reduction in isometric load under high [Pi], manifested as a downward shift of the force – velocity curve (Figs. 6a-b). This effect is especially pronounced at low stimulation frequencies. For example, at 40 Hz, the isometric load of a motor unit with 50 fibers drops from about $0.45T_1$ at 1 mM [Pi] to approximately $0.25T_1$, while a unit with 100 fibers shows a decrease from $0.9T_1$ to roughly $0.5T_1$.

The downward shift of the force-velocity curve would lead to an overall reduction in output power, indicating impaired muscle performance. However, elevated [Pi] also reduces ATP consumption (Figs. 6c-d). Our simulations suggest that cross-bridge efficiency increases under high [Pi]: the peak efficiency rises from around 30% (Fig. 4e,f) to be close to or even over 40% (Figs. 6e,f). This occurs mainly because elevated [Pi] slows the release of phosphate from myosin in the model, thereby improving efficiency (20).

The shape of the efficiency curve—rising then falling with load—remains unchanged under high [Pi]. The previously observed size-dependent efficiency pattern also persists: at low loads, smaller units show higher efficiency, whereas at high loads, larger units become more efficient. This holds across both low and high stimulation frequencies. A notable difference under elevated [Pi] is that the efficiency gap between different stimulation frequencies widens. At low [Pi], the efficiency difference between 40 Hz and 100 Hz was less than 5% in Fig. 4; at 20 mM [Pi], this difference increases to be over 10% in Fig. 6.

The amplified efficiency gap between high and low stimulation frequencies under elevated [Pi] (20 mM) in our simulations is primarily driven by the slowing of Pi release due to increased Pi rebinding. The mechanism can operate as follows: elevated [Pi] increases Pi rebinding, which prolongs the lifetime of the AM*ADP·Pi state and allows myosin motors to undergo repeated detachment-rebinding cycles without consuming additional ATP. This “energy reuse” mechanism enhances cross-bridge efficiency, and the effect is more pronounced at higher stimulation frequencies when more motors are active, while $Ca^{2+}$ precipitation plays a negligible role under our conditions.

Since it was reported that 20 mM [Pi] lowers myoplasmic [$Ca^{2+}$] by about 29% due to precipitation in the SR (63), we tested the potential confounding effect of reduced cytosolic [$Ca^{2+}$] and performed additional simulations at 100 Hz stimulation in which we deliberately reduced BMP [$Ca^{2+}$] by 29% while keeping all other parameters unchanged. The resulting efficiency gap between 40 Hz and 100 Hz in our simulations slightly increases further, as seen from Fig. S6 in Supplemental Information. This suggests that reduced $Ca^{2+}$ availability due to precipitation in the SR may also have some effect on the efficiency gap under high [Pi].

**Discussion**

Our model predictions suggest that the emergent shortening velocity—resulting from the coordinated movement of multiple motors during muscle contraction—is the primary determinant of cross-bridge efficiency. At low shortening velocities, motors on average experience high loads (Fig. 5c) but small swing distances (Fig. 5b), and the fraction of motors that complete a full power stroke decreases (Fig. 5a). At high shortening velocities, motors on average experience large swing distances but low loads, which again lowers efficiency. Only at intermediate shortening velocities do motors on average experience both high loads and large swing distances. This leads to the bell-shaped efficiency-velocity relationship in our simulation.

In our model, stimulation frequency alters the number of working motors via calcium-dependent activation, which would change internal strain distribution along the thin filament due to coordinated movement among multiple motors. Because several kinetic rates of myosin motors are force-sensitive in our model, the efficiency-velocity relationship could in principle shift with the number of working motors. Therefore, the fact that our simulation results show cross-bridge efficiency is primarily determined by the emergent shortening velocity (Fig. 3e) is not a built-in outcome, even though cross-bridge kinetic rates in our model are not directly modulated by stimulation frequency.

In fact, we can define the causal pathway in our model as follows: Stimulation

Frequency → Activation Level → Contractile Dynamics → Efficiency. As shown in Fig. 2b, stimulation frequency directly modulates $R_{active}$. For a given filament load, a change in $R_{active}$ alters the shortening velocity in Fig. 3a. In Fig. 3e, we find that the velocity shift accounts for the majority of the efficiency variation observed across frequencies in our simulations. We therefore characterize shortening velocity as the "primary determinant" of cross-bridge efficiency. Meanwhile, as seen from Fig. 3e, frequency appears to exert a secondary, velocity-independent influence on efficiency. A lower $R_{active}$ at low frequencies can lower the cross-bridge work-to-ATP ratio even when shortening velocity is matched.

Though a direct quantitative comparison between our simulations and experimental observations is challenging due to species and protocol differences across different experimental studies (12-15, 61), we have made efforts to assess the model's predictive capability on some key performance metrics. For example, our model predicts a peak cross-bridge efficiency of approximately 30% at 1 mM [Pi], which is close to the previous reported values in experiments (64-66). The model also predicts that peak efficiency occurs at a shortening velocity of VB ≈ 0.25, closely matching the experimental range of approximately 0.2 (6, 10). It is worth noting that the efficiency values from our model are higher than those typically measured in whole-muscle experiments. This discrepancy arises mainly because our model focuses exclusively on cross-bridge efficiency and does not include the energy costs associated with the activation heat.

It is worth noting that Buschman et al. (13, 61) measured the mechanical efficiency of muscle fibers at different stimulation frequencies. By subtracting the heat produced during activation without shortening, they derived the cross-bridge efficiency, which was found to decrease at higher stimulation frequencies (13, 61). Specifically, Buschman et al. (13, 61) reported that cross-bridge efficiency was high at low activation levels in isolated single Xenopus muscle fibers, where recruitment confounders are absent. We suspect that the discrepancy between their findings (decreasing efficiency with stimulation frequency) and our model's predictions (increasing efficiency with stimulation frequency) likely arises from several factors. In Buschman et al. (13, 61),

“cross-bridge efficiency” was not measured directly; instead, it was derived by subtracting the activation heat measured under isometric conditions from the total energy released during shortening. This approach assumes that the energy cost of activation processes, such as calcium pumping, remains independent of shortening. However, it was suggested that shortening can itself modulate activation processes to generate additional shortening heat (67). Moreover, the calculated cross-bridge efficiency reported by Buschman et al. appeared to be very high—up to 50% (61) or 80% (13). It has been suspected that a portion of pre-stored elastic energy might be released during unfused contraction at low stimulation frequencies (14), which could have contributed to the reported high efficiency. In contrast, our model calculates intrinsic cross-bridge efficiency, providing a “purer” mechanical metric that does not include activation-related metabolic noise.

It is also noteworthy that Buschman et al. (13, 61) used fibers from Xenopus laevis, which may exhibit different thermal and kinetic properties compared to mammalian fibers. Conversely, studies on mammalian skeletal muscle (14) reported that cross-bridge efficiency increases with activation level. These contradictory results might suggest that the relationship between stimulation frequency and efficiency is species-dependent, which, however, have not been thoroughly investigated in our current work due to the lack of species-specific parameters.

In the end, we must emphasize the limitations of our current model. Some limitations related to sarcomere mechanics were already listed in our previous work (20). In addition, the fine activation mode of thin filament (68) has not been considered in the model yet. We also acknowledge that the parameters used in our simulations are specific to the conditions studied, which should be kept in mind when interpreting our findings.

**Conclusion**

By integrating calcium-mediated excitation with a detailed cross-bridge cycle, our biophysical model reveals that skeletal muscle efficiency is governed primarily by shortening velocity, with efficiency following a bell-shaped dependence on velocity that peaks at a near-constant optimal value across frequencies. Stimulation frequency exerts only a secondary influence, modulating efficiency largely through its regulation of shortening velocity, which in turn determines the power-stroke characteristics of myosin motors. This velocity-centric mechanism explains why peak efficiency remains remarkably consistent across a range of stimulation frequencies, with only a modest shift toward higher velocities at elevated frequencies. Such a mechanism may enable muscles to contract at their optimal speed with maximal efficiency, owing to strain compatibility across multiple motor units within a single muscle. Furthermore, we find that elevated inorganic phosphate ([Pi]) amplifies the efficiency gap between high- and low-frequency conditions. Together, these findings may help resolve apparent contradictions in prior experimental literature and establish a unified, physiologically grounded framework for understanding how neural drive—via its control of contraction dynamics—shapes muscle energetics in both health and disease.

## Acknowledgements

This work was supported by the National Natural Science Foundation of China (Grant No.: 12372318) and Zhejiang Provincial Natural Science Foundation of China (Grant No.: LZ23A020004).

## Data Availability

Data will be shared upon reasonable request.

## Author contributions

B.C. designed research, J.X. and B.C. performed research, J.X. and B.C. analyzed data, J.X. and B.C. wrote the manuscript.

## Declaration of Interest

The authors declare no competing interests.

**Supplemental Information**

a) Model validation

We have performed a series of baseline simulations to validate the feasibility of the model against previously published experimental data. First, the response of a muscle fiber to a single action potential is simulated. As shown in Fig. S1a, force peaks at approximately 42 ms, with a half-relaxation time ($t_{50}$) of about 18 ms, which is comparable to experimental measurements (69). After the peak, force gradually declines, returning near the pre-stimulation isometric level after roughly 250 ms, which is also comparable to the experimental result (69).

We have furtherly examined a tetanic contraction by simulating the effect of membrane potential repolarization to rest following a period of high-frequency stimulation (Fig. S1b). Beginning from a stable resting baseline, a 200 Hz stimulus train is applied at 0.05 s, eliciting force development. At 0.2 s, the membrane potential is returned to rest. The force relaxation rate, defined as the time for force to decay to half its value at stimulus cessation, is approximately 36 ms - again comparable to experimental observations (60, 69).

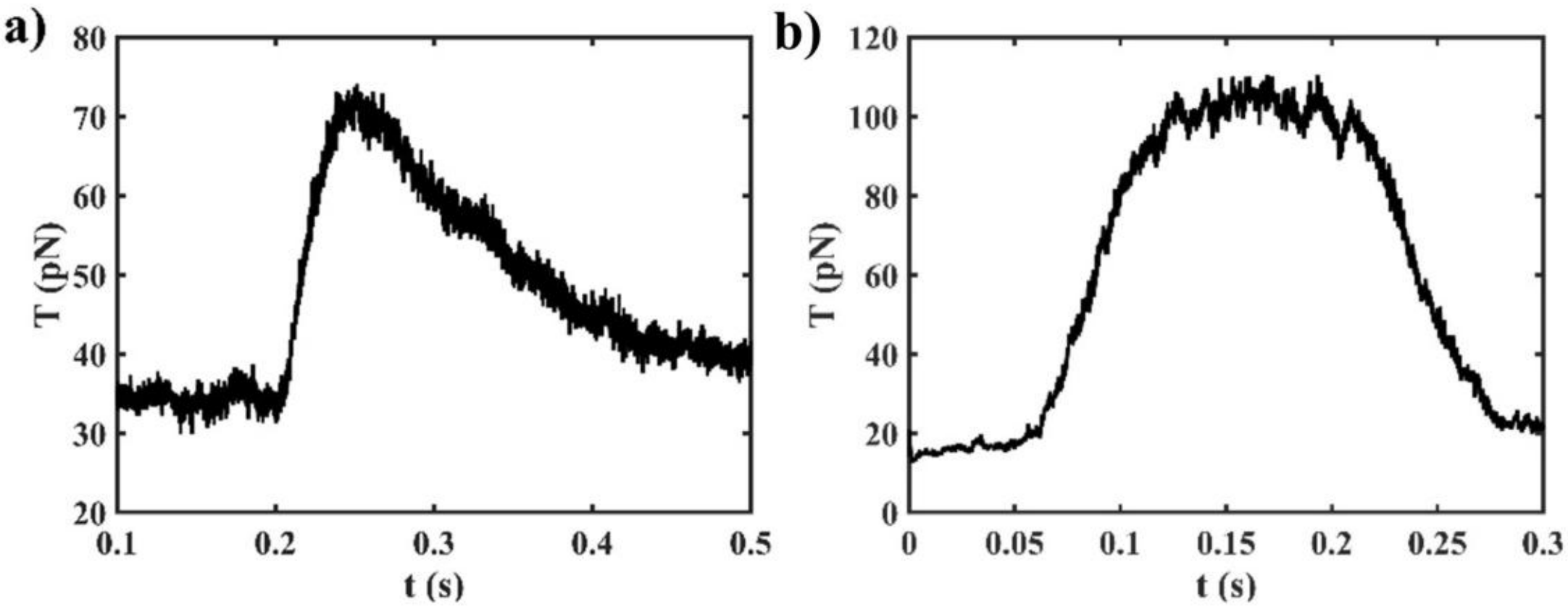

**Fig. S1** a) Evolution of muscle fiber tension over time during a single twitch; b) Evolution of muscle fiber tension over time during a period of tetanic contraction.

b) Error Bar

In Monte Carlo simulations, to ensure that our conclusions are robust and not artifacts of random fluctuations, each simulation is repeated 100 times, and the results are averaged to generate the data presented in the main figures. The error analysis for selected data from Fig. 3 is presented in Fig. S2, where only the results for stimulation frequencies of 40 Hz and 100 Hz are displayed. The results indicate that the variability of our data remains within an acceptable range and should not affect the validity of the conclusions.

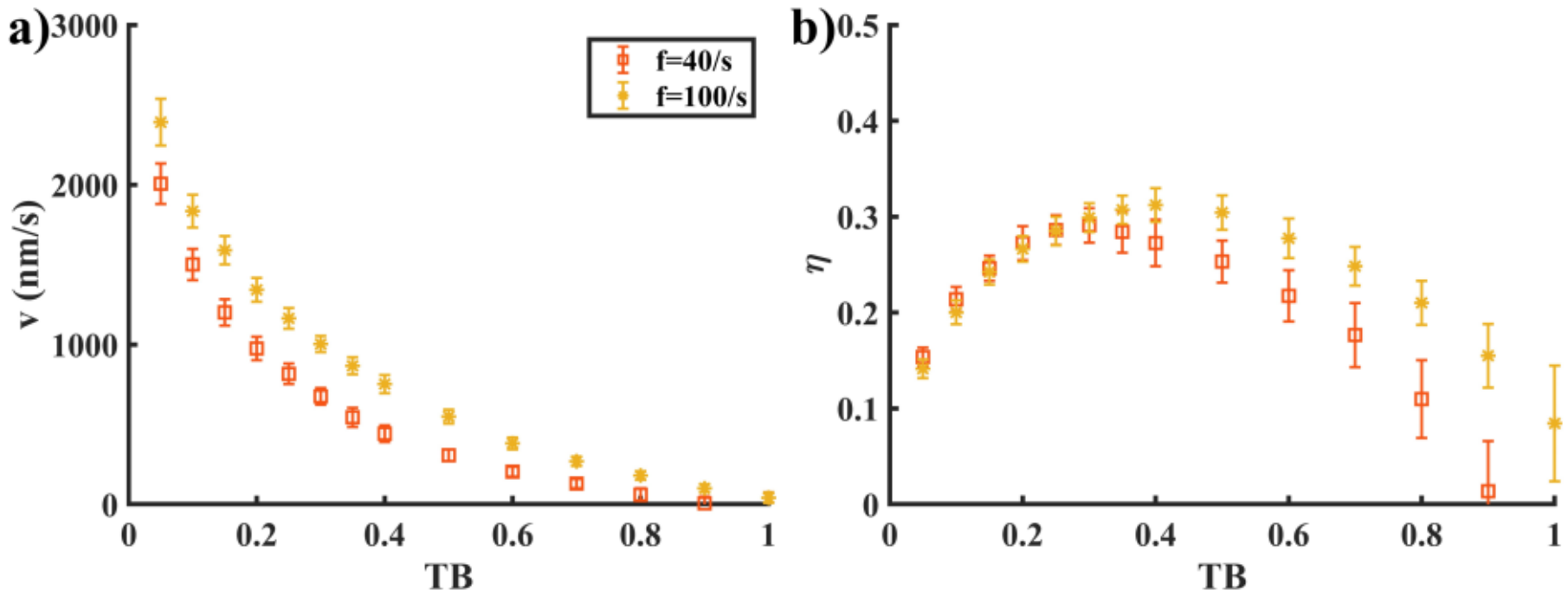

**Fig. S2** a) Error bar of the shortening velocity; b) Error bar of the energy efficiency. In the simulation, *f*=40/s and 100/s.

c) Chemical reaction equations

Chemical reactions incorporated into the model are provided in the following (36).

Reaction (1) is for the binding of $Ca^{2+}$ to ATP, forming Ca-ATP, where the forward and reverse reaction rates are denoted as $v_{(CaATP)+}$ and $v_{(CaATP)-}$, respectively.

$$Ca^{2+} + ATP \underset{k_{(CaATP)-}}{\overset{k_{(CaATP)+}}{\rightleftharpoons}} CaATP \tag{1}$$

$$v_{(CaATP)+} = k_{(CaATP)+}[Ca^{2+}][ATP];$$

$$v_{(CaATP)-} = k_{(CaATP)-}[CaATP]$$

Reaction (2) is for the binding of $Mg^{2+}$ to ATP, forming Mg-ATP, where the forward and reverse reaction rates are denoted as $v_{(MgATP)+}$ and $v_{(MgATP)-}$, respectively.

$$Mg^{2+} + ATP \underset{k_{(MgATP)-}}{\overset{k_{(MgATP)+}}{\rightleftharpoons}} MgATP \tag{2}$$

$$v_{(MgATP)+} = k_{(MgATP)+}[Mg^{2+}][ATP];$$

$$v_{(MgATP)-} = k_{(MgATP)-}[MgATP]$$

Reaction (3) is for the binding of $Ca^{2+}$ to parvalbumin (abbreviated as Pa) in the myoplasm following its release, forming Pa-Ca, where the forward and reverse reaction rates are denoted as $v_{(Pa-Ca)+}$ and $v_{(Pa-Ca)-}$, respectively.

$$Ca^{2+} + Pa \underset{k_{(Pa-Ca)-}}{\overset{k_{(Pa-Ca)+}}{\rightleftharpoons}} Pa - Ca \tag{3}$$

$$v_{(Pa-Ca)+} = k_{(Pa-Ca)+}[Ca^{2+}][Pa];$$

$$v_{(Pa-Ca)-} = k_{(Pa-Ca)-}[Pa - Ca]$$

Reaction (4) is for the binding of $Mg^{2+}$ to parvalbumin, forming Pa-Mg, where the forward and reverse reaction rates are denoted as $v_{(Pa-Mg)+}$ and $v_{(Pa-Mg)-}$, respectively.

$$Mg^{2+} + Pa \underset{k_{(Pa-Mg)-}}{\overset{k_{(Pa-Mg)+}}{\rightleftharpoons}} Pa - Mg \tag{4}$$

$$v_{(Pa-Mg)+} = k_{(Pa-Mg)+}[Mg^{2+}][Pa];$$

$$v_{(Pa-Mg)-} = k_{(Pa-Mg)-}[Pa - Mg]$$

Reaction (5) is for the binding of $Ca^{2+}$ to calsequestrin (abbreviated as Cs) within the sarcoplasmic reticulum (SR), forming Cs-Ca, where the forward and reverse reaction rates are denoted as $v_{(Cs-Ca)+}$ and $v_{(Cs-Ca)-}$, respectively.

$$Ca^{2+} + Cs \underset{k_{(Cs-Ca)-}}{\overset{k_{(Cs-Ca)+}}{\rightleftharpoons}} Cs - Ca \tag{5}$$

$$v_{(Cs-Ca)+} = k_{(Cs-Ca)+}[Ca^{2+}][Cs];$$

$$v_{(Cs-Ca)-} = k_{(Cs-Ca)-}[Cs - Ca]$$

Reactions (6-7) describe the sequential binding of two $Ca^{2+}$ ions to troponin (abbreviated as Tn). The first step involves Tn binding one $Ca^{2+}$ to form Tn-Ca, with forward and reverse rates being denoted as $v_{(Tn-Ca)+}$ and $v_{(Tn-Ca)-}$, respectively.

$$Ca^{2+} + Tn \underset{k_{Ca-}}{\overset{k_{Ca+}}{\rightleftharpoons}} Tn - Ca \tag{6}$$

$$v_{(Tn-Ca)+} = k_{Ca+}[Ca^{2+}][Tn];$$

$$v_{(Tn-Ca)-} = k_{Ca-}[Tn - Ca]$$

The second step involves Tn-Ca binding a second $Ca^{2+}$ to form Tn-2Ca, with forward and reverse rates being denoted as $v_{(Tn-2Ca)+}$ and $v_{(Tn-2Ca)-}$, respectively.

$$Ca^{2+} + Tn - Ca \underset{k_{Ca-}}{\overset{k_{Ca+}}{\rightleftharpoons}} Tn - 2Ca \quad (7)$$

$$v_{(Tn-2Ca)+} = k_{Ca+}[Ca^{2+}][Tn - Ca];$$

$$v_{(Tn-2Ca)-} = k_{Ca-}[Tn - 2Ca]$$

The reaction rate constants used in the aforementioned reactions are provided in Table S1, together with the initial concentrations of reactants involved in the model. The parameters are defined as follows: $[Pa]_0$ represents the initial concentration of parvalbumin in the myoplasm; $[Cs]_0$, the initial concentration of calsequestrin in the SR; $[Tn]_0$, the initial concentration of troponin in the myoplasm; $[ATP]_0$, the initial concentration of ATP in the myoplasm; and $[Mg]_0$, the initial concentration of $Mg^{2+}$ in the myoplasm. The initial concentration of $Ca^{2+}$ is higher within the SR, denoted as $[Ca_{SR}]_0$, and lower in the myoplasm, denoted as $[Ca_{MP}]_0$. It is important to note that, although the SR region is subdivided into GSR and TSR compartments, and the myoplasm into BMP and TMP compartments, the initial concentrations of $Ca^{2+}$ and calsequestrin are assumed to be identical in GSR and TSR. Similarly, the initial concentrations of various ions and organic molecules are set to be equal in the BMP and TMP compartments (21).

**Table S1** Default values of parameters for chemical reaction equations used in the simulations

| Parameter | Value | Parameter | Value |
|---|---|---|---|
| $k_{(CaATP)+}$ | 0.15 /$\mu M$ms (36) | $k_{(CaATP)-}$ | 30 /ms (36) |
| $k_{(MgATP)+}$ | 0.0015 /$\mu M$ms (36) | $k_{(MgATP)-}$ | 0.15 /ms (36) |
| $k_{(Pa-Ca)+}$ | 0 /$\mu M$ms (36) | $k_{(Pa-Ca)-}$ | 0 /ms (36) |
| $k_{(Pa-Mg)+}$ | 0 /$\mu M$ms (36) | $k_{(Pa-Mg)-}$ | 0 /ms (36) |
| $k_{(Cs-Ca)+}$ | 0.000012 /$\mu M$ms (35) | $k_{(Cs-Ca)+}$ | 0.01 /ms (35) |
| $k_{Ca+}$ | 0.0885 /$\mu M$ms (36) | $k_{Ca-}$ | 0.115 /ms (36) |
| $[Pa]_0$ | 1500 $\mu M$ (36) | $[Cs]_0$ | 31000 $\mu M$ (35) |
| $[Ca_{SR}]_0$ | 1480 $\mu M$ (21) | $[Ca_{MP}]_0$ | 50 $\mu M$ (21) |
| $[ATP]_0$ | 2000 $\mu M$ | $[Mg]_0$ | 2000 $\mu M$ |
| $[Tn]_0$ | 140 $\mu M$ (21) | | |

d) Rate formula for some processes in the model

The potential of a single twitch in the model is approximated as a piecewise exponential function: $V = \begin{cases} 115\text{mV} \exp(6(\text{t}-0.5)) - 80\text{mV} & t \le 0.5ms \\ 115\text{mV} \exp(-1.5(\text{t}-0.5)) - 80\text{mV} & t > 0.5ms \end{cases}$. The rate for the transition of each voltage sensor of DHPR from the resting state to the activated state is given by $k_D = 0.5\alpha \exp\left(\frac{V-\bar{V}}{8\text{K}}\right)$. The rate for the transition of each voltage sensor of DHPR from the activated state to the resting state is given by $k_{-D} = 0.5\alpha \exp\left(-\frac{V-\bar{V}}{8K}\right)$ (31). The binding rate of Myosin and actin is given by $k_{bi} = \sqrt{\frac{\beta}{\pi}} 2\xi \exp(-\beta U^2) / [1 + \text{erf}(0.3\sqrt{\beta})]$, where $\beta = q l_{\text{a}}^2/(2k_{\text{B}}T)$ and $U = u/l_{\text{a}}$, $k_{\text{B}}$ is the Boltzmann constant, and $T$ is the absolute temperature (20). Catch-slip bond rates are given by

$$k_{break1} = \begin{cases} 166/s \exp\left(-\frac{f}{3}\right) + 3.32/s \exp\left(\frac{f}{1.2}\right) & f \le 5pN \\ 178.45/\text{s} + 166/s \exp\left(-\frac{f}{3}\right) + 12.45/s \exp\left(\frac{f}{5}\right) & f > 5pN \end{cases}$$

and $k_{break2} = 199/s \exp\left(-\frac{f}{1.5}\right) + 8.3/s \exp\left(\frac{f}{5}\right)$, respectively (20).

Table S2 Default values of parameters for rate formulas used in the simulations

| Parameter | Value | Parameter | Value |
|---|---|---|---|
| $K$ | 4.5 mV (31) | $\alpha$ | 0.2 /ms (31) |
| $\bar{V}$ | -20 mV (31) | $k_{\text{B}}T$ | $4.14\times10^{-21}$ J |

e) Parameter sensitivity analysis

The sensitivity analysis on some key parameters, including $\gamma$, $k_{-R}$, and $k_{\mathrm{a}}$ are listed in Figs. S3-S5, where each parameter was scaled to 80%, 90%, 100% (baseline), 110%, and 120% of its original value to investigate its impact on the performance of a single muscle fiber. Among these parameters, the model exhibits a relatively high degree of sensitivity to $\gamma$.

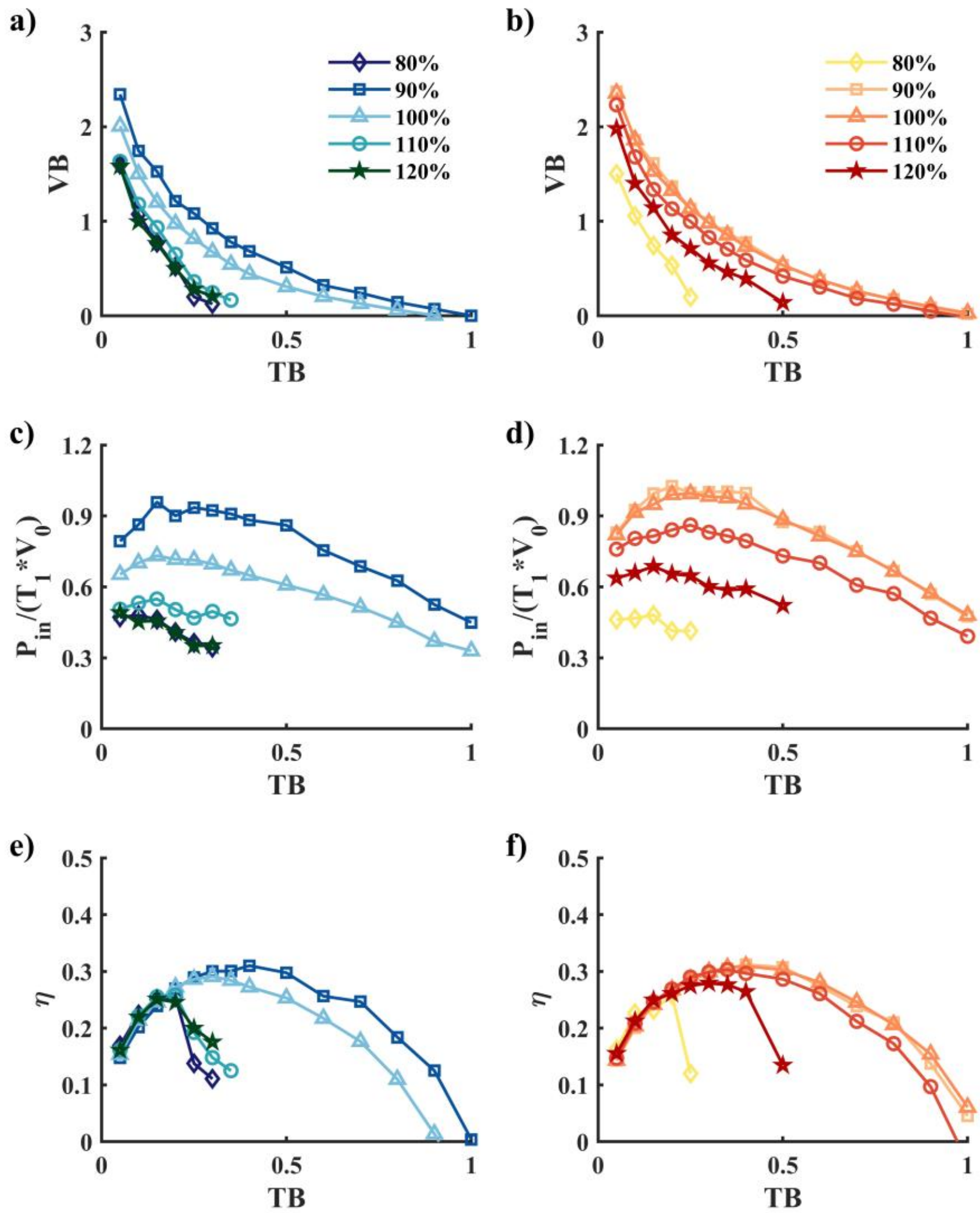

**Fig. S3** Sensitivity analysis of $\gamma$ with a stimulation frequency of 40Hz (a, c, e) and 100Hz (b, d, f), where $\gamma$ was scaled to 80%, 90%, 100%, 110%, and 120% of its baseline value: (a, b) Shortening velocity vs. load; (c, d) Power consumption vs. load; (e, f) Energy efficiency vs. load.

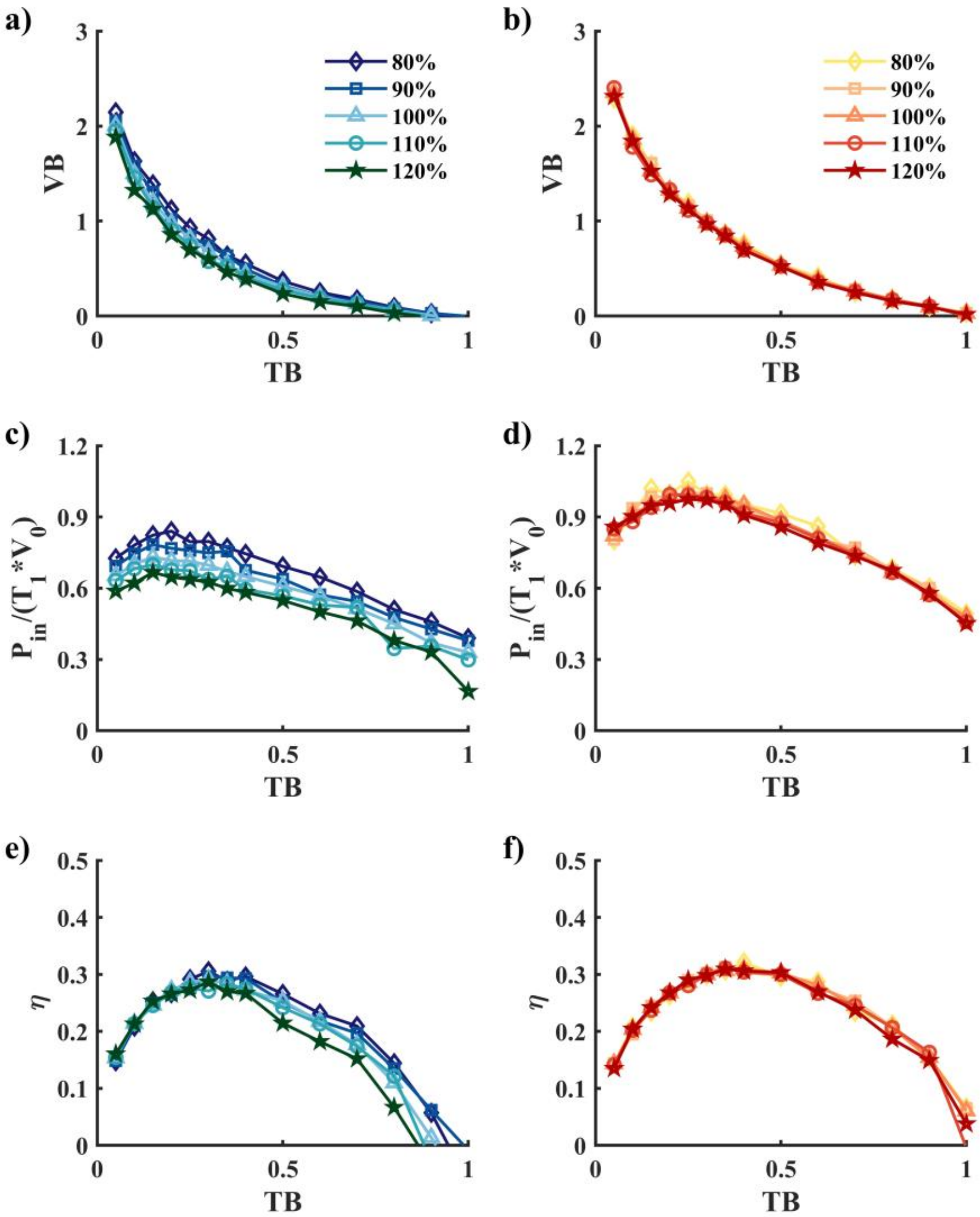

**Fig. S4** Sensitivity analysis of $k_{-R}$ with a stimulation frequency of 40Hz (a, c, e) and 100Hz (b, d, f) where $k_{-R}$ was scaled to 80%, 90%, 100%, 110%, and 120% of its baseline value: (a, b) Shortening velocity vs. load; (c, d) Power consumption vs. load; (e, f) Energy efficiency vs. load.

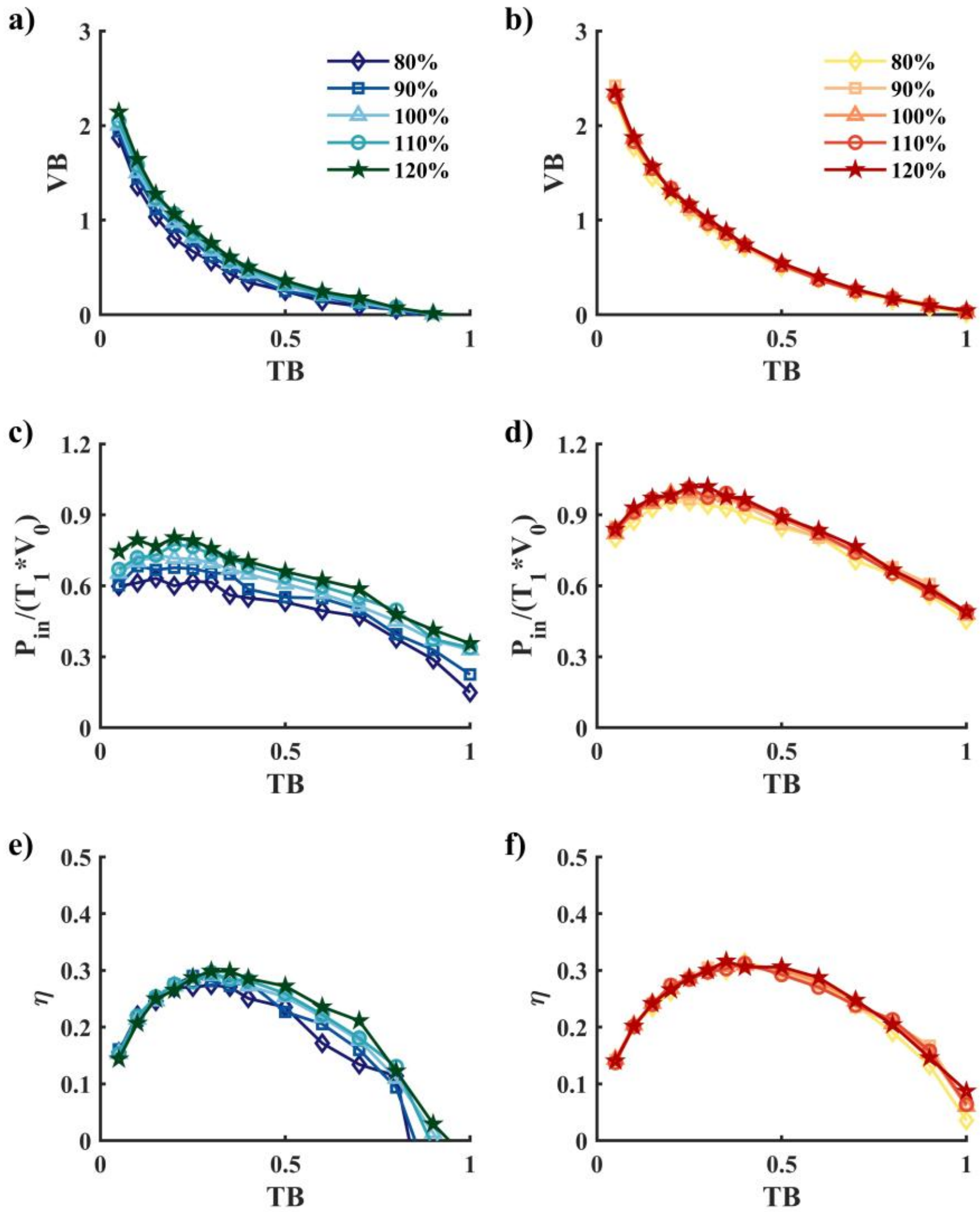

**Fig. S5** Sensitivity analysis of $k_{\mathrm{a}}$ with a stimulation frequency of 40Hz (a, c, e) and 100Hz (b, d, f)where $k_{\mathrm{a}}$ was scaled to 80%, 90%, 100%, 110%, and 120% of its baseline value: (a, b) Shortening velocity vs. load; (c, d) Power consumption vs. load; (e, f) Energy efficiency vs. load.

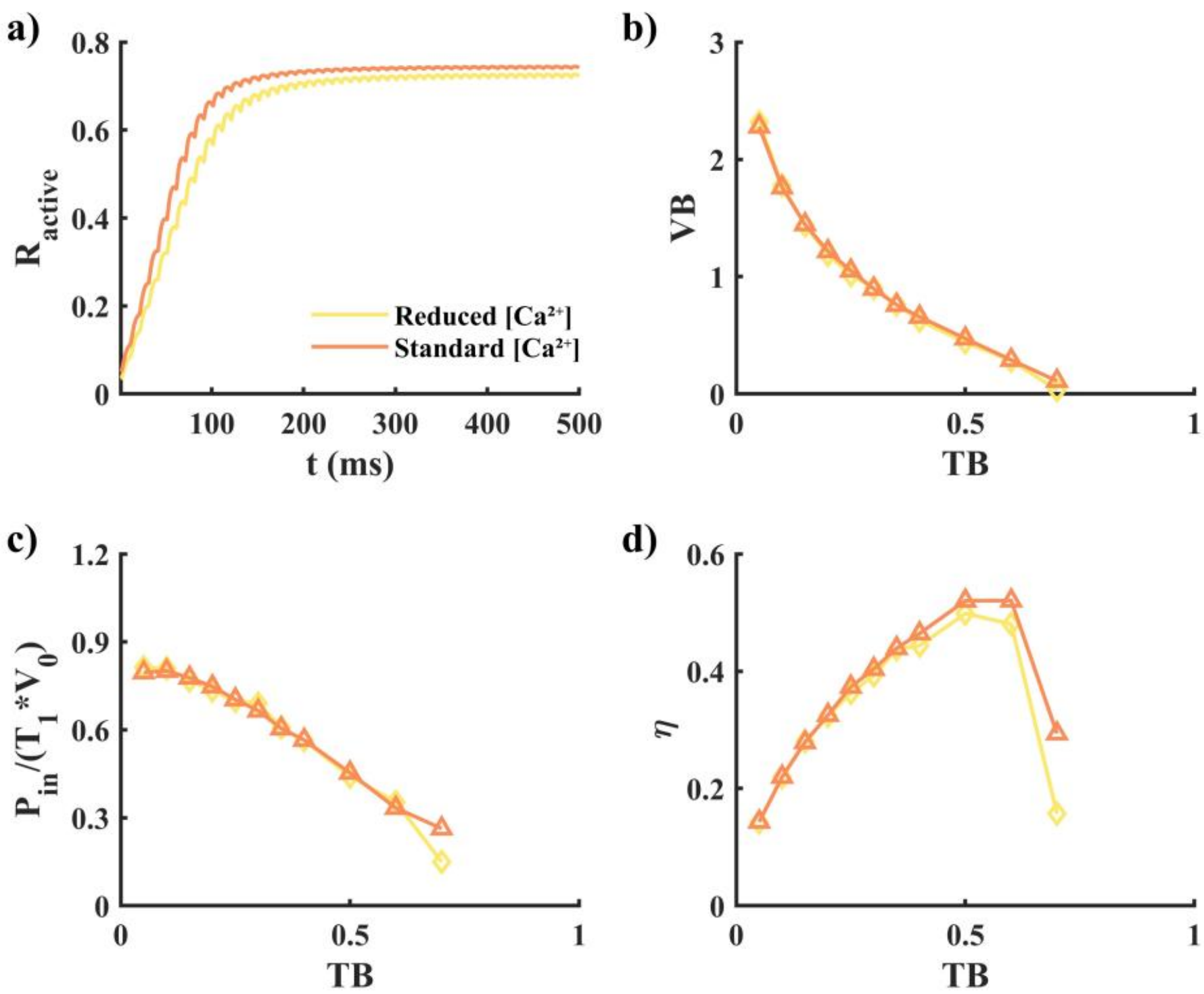

**Fig. S6** Comparison of muscle contraction features between the reduced [$Ca^{2+}$] case simulating a 29% decrease due to Ca-Pi precipitation and the standard [$Ca^{2+}$] case without precipitation under [Pi] = 20 mM. Results are shown for stimulation frequencies of 100 Hz: (a) Time evolution of $R_{active}$ on the thin filament; (b) Shortening velocity vs. load; (c) Power consumption vs. load; (d) Energy efficiency vs. load.